# Seeking Spectroscopic Binaries with Data-driven Models

Isabel Angelo[1], Erik Petigura[1], and Megan Bedell[2]
[1] Department of Physics & Astronomy, University of California Los Angeles, Los Angeles, CA 90095, USA
[2] Center for Computational Astrophysics, Flatiron Institute, 162 5th Avenue, New York, NY 10010, USA

## Abstract

Data-driven stellar classification has a long and important history in astronomy, dating as far back as Annie Jump Cannon's "by-eye" classifications of stars into spectral types, still used today. In recent years, data-driven spectroscopy has proven to be an effective means of deriving stellar properties for large samples of stars, sidestepping issues with computational efficiency, incomplete line lists, and radiative transfer calculations associated with physical stellar models. A logical application of these algorithms is the detection of unresolved stellar binaries, which requires accurate spectroscopic models to resolve flux contributions from a fainter secondary star in the spectrum. Here, we use The Cannon to train a data-driven model on spectra from the Keck High Resolution Echelle Spectrometer. We show that our model is competitive with existing data-driven models in its ability to predict stellar properties $T_{\rm eff}$, $R_\star$, [Fe/H], and $v \sin i$, as well as the instrumental point-spread function, particularly when we apply a novel wavelet-based processing step to spectra before training. We find that even with accurate estimates of star properties, our model's ability to detect unresolved binaries is limited by its ∼3% accuracy in per-pixel flux predictions, illuminating possible limitations of data-driven model applications.

## 1. Introduction

Across all subfields of astronomy, the characterization of astronomical objects hinges on stellar spectroscopy. Spectroscopic observations of stars enable us to precisely determine their age, rotation, and chemical abundances, which play an important role in understanding the formation and evolution of objects as small as planets and as large as galaxies. In general, these properties are derived via comparisons of stellar spectra to synthetic spectra derived from numerical approximations of stellar structure and radiative transfer (e.g., R. L. Kurucz 1979; T. O. Husser et al. 2013). However, these models do not always resemble true observations of stellar spectra, due to incomplete atomic line lists and numerous simplifications of the underlying physics needed to achieve sufficient computational efficiency.

In recent years, several studies have successfully used machine learning to address these issues, particularly in the context of determining "stellar labels" or properties and abundances that can be derived from stellar spectra. Simple linear regression models of stellar flux as a function of stellar labels have made consistent and accurate predictions of stellar properties like $T_{\rm eff}$, $\log g$, [Fe/H], and $v \sin i$ (e.g., The Cannon—M. Ness et al. 2015). Additionally, more complex neural networks are also viable means of determining stellar properties from stellar spectra (e.g., The Payne—Y.-S. Ting et al. 2019). These applications are particularly successful for large-scale surveys, where ample training data exist and high-fidelity labels of previous surveys can be leveraged to train the model. Indeed, data-driven models have been trained to predict stellar labels for spectra for a number of surveys, including APOGEE (M. Ness et al. 2015), LAMOST (A. Y. Q. Ho et al. 2017), and Gaia (I. Angelo et al. 2024).

Data-driven spectroscopic models also pose an exciting opportunity for science cases that require accurate predictions of the spectra themselves. For example, previous attempts to identify unresolved binaries in stellar spectra have historically been limited to close binaries (orbital separation $<10$ au) with large radial velocity offsets that produce two distinct sets of spectral features. However, in recent years, accurate data-driven spectral models have been used to identify thousands of unresolved binaries with $>10$ au separations, which manifest as overlapping spectral features that produce subtle deviations from single-star spectra. In particular, K. El-Badry et al. (2018a, 2018b) developed a data-driven spectral model that identified over 3000 unresolved binaries in APOGEE. More recently, similar searches have been carried out with synthetic template spectra, identifying signatures of unresolved binaries in larger surveys like Gaia-ESO (M. Kovalev & I. Straumit 2022) and LAMOST (M. Kovalev et al. 2024) and in targeted ground-based observations of individual objects (J. F. González et al. 2024).

Applying these techniques to identify unresolved binaries in the planet host population is of particular interest. A large majority of the planets discovered to date come from space-based photometric missions like Kepler, K2, and TESS. However, these missions' large $\geqslant 4''$ pixels and considerable distances to most targets mean that many known planet hosts may have additional unresolved companions. Furthermore, follow-up studies to identify stellar companions to planet hosts have largely relied on ground-based adaptive optics imaging, which is only sensitive to companions beyond $>50$ au for most Kepler stars.

The California-Kepler Survey (CKS) provides an exciting opportunity to search for $<50$ au binaries in the planet census. The CKS survey is a Kepler follow-up program that obtained high-resolution spectra of over a thousand planet hosts using the High Resolution Echelle Spectrometer (HIRES; S. S. Vogt et al. 1994) instrument on Keck I, providing a large sample of targets to search for binaries. Additionally, ample training data of well-characterized stars and high-fidelity stellar labels are available. In fact, HIRES spectra have already been used to develop multiple data-driven spectroscopic models (S. W. Yee et al. 2017; M. Rice & J. M. Brewer 2020) that accurately predict properties of stars with HIRES spectra. Finally, the high signal-to-noise ratio (S/N) of HIRES spectra offers the ability to detect the subtle signatures that an unresolved secondary star imprints on the stellar spectrum.

In this paper, we explore potential applications of data-driven models to identify unresolved binaries in the CKS sample. In Section 2, we use The Cannon—a data-driven framework for modeling stellar spectra—to develop a spectroscopic model of single-star and binary HIRES spectra. We also develop a novel wavelet-based filtering step to treat the stellar continuum and show that it significantly improves our model's ability to accurately predict stellar properties (Section 2.2.3). We evaluate our model's ability to determine the properties of single stars and identify unresolved binaries in Sections 3 and 4, respectively. We find that our data-driven model does not confidently resolve single-star and binary populations in our validation data, likely due to its limited accuracy in predicting the per-pixel flux of the single-star spectra in our sample. Finally, in Section 5, we present simple experiments with synthetic HIRES spectra to contextualize our findings and show that unresolved binaries across a wide range of parameter space should produce signatures of binarity in HIRES spectra. We also comment on the implications of our findings for future applications and limitations of data-driven spectroscopy.

## 2. A Data-driven Spectral Binary Emulator

We are particularly interested in identifying binaries with overlapping spectral features ("spectral binaries") that would have been missed by more traditional searches for SB2s binaries with multiple sets of spectral features (SB2s). The concept of spectral binaries was first introduced in the context of hot stars with cooler or substellar companions (A. J. Burgasser 2007; D. C. Bardalez Gagliuffi et al. 2014), but here we refer more generally to binaries with overlapping spectral features and components of different spectral types. This includes binaries with smaller radial velocity offsets between the spectral features of the primary and secondary star, either due to wider (>10 au) separations or smaller orbital inclinations. We can identify these binaries as spectra that are significantly better fit by a composite spectrum—that is, a linear combination of two single-star spectra—than by a single-star spectrum (K. El-Badry et al. 2018a). This requires an accurate model of single-star spectra that can be used to construct a composite spectrum model.

Historically, single-star spectra have typically been modeled using ab initio stellar models, which are derived from numerical approximations of stellar structure and radiative transfer. We experimented with using ab initio stellar models from Starfish (I. Czekala et al. 2015) to model single-star spectra; however, we found that these models disagreed in many cases with the per-pixel flux of real HIRES spectra by 10%–20%. This is much larger than the <10% flux contributions we expect from secondary stars in spectral binaries (see Section 5.1 for a derivation of this prediction), thus these ab initio models are not feasible for our purposes.

In recent years, stellar spectra have been increasingly modeled by data-driven models, which are trained on large samples of stellar spectra to accurately predict star properties based on their spectra (see, for example, M. Ness et al. 2015; A. Y. Q. Ho et al. 2017; A. Behmard et al. 2019). While these models have typically been used to predict star properties like $T_{\rm eff}$, $\log g$, and chemical abundances, the fact that they are trained on real spectra suggests that they may also be able to accurately predict spectra of stars observed by the survey that they are trained on. Thus, we investigated whether a data-driven spectral emulator trained on HIRES spectra can predict single-star spectra with sufficient accuracy to identify signatures of spectral binaries.

### 2.1. Training Data Selection

We trained our data-driven model on a sample of high-S/N HIRES spectra of well-characterized stars from the SpecMatch-Emp library. This library was published in S. W. Yee et al. (2017) and consists of 404 bright, well-characterized stars spanning a range of spectral types, ~F1–M5. We selected this sample as our model training set due to its uniform coverage of stellar label space and high average per-pixel S/N $\sim$ 150. We then removed any reported unresolved binaries from our training set, since unresolved binaries will manifest as biased stellar parameters in our model training set (K. El-Badry et al. 2018a). To do this, we used the SIMBAD Astronomical Database to query the SpecMatch-Emp library and remove any unresolved binaries from our training set. In particular, we removed 22 stars flagged in SIMBAD as spectroscopic, eclipsing, or X-ray binaries and 10 additional stars with reported stellar companions within $1\farcs5$. We also removed eight stars flagged by Gaia's `non_single_star` metric and an additional 29 stars with reported renormalized unit weight error (see V. Belokurov et al. 2020) > 1.4. We then divided our sample into hot and cool samples to train separate hot- and cool star models, as described further in Section 3.

A number of SpecMatch-Emp library stars do not have reported $v \sin i$ values. We removed these stars from our hot-star sample, but because they comprised such a large fraction of the cool star sample (≈55%), we could not remove them without compromising our model performance. In the cool star case, the $v \sin i$ values are typically not reported, due to challenges in determining the $v \sin i$ values for slow rotators ($\lesssim$2 km s$^{-1}$), thus we can assume the $v \sin i$ values for these stars $\leqslant$2 km s$^{-1}$. In order to preserve the information from these spectra and their associated labels, we included these spectra with assigned $v \sin i = 0$ km s$^{-1}$ in our training set, along with "broadened" copies of each spectrum that we broadened to $v \sin i = 3$, 5, and 7 km s$^{-1}$ with a rotational broadening kernel (Equation (17.12) in D. F. Gray 1992). Our final training set is composed of 123 unique hot stars, 128 unique cool stars, and 291 broadened cool stars. The distribution of training labels in our models is shown in Figure 1, and the spectroscopic properties of our training set are summarized in Table 1.

### 2.2. Data Preparation

The Cannon requires rest-frame, continuum-normalized spectra as inputs for its training step. The HIRES spectra that

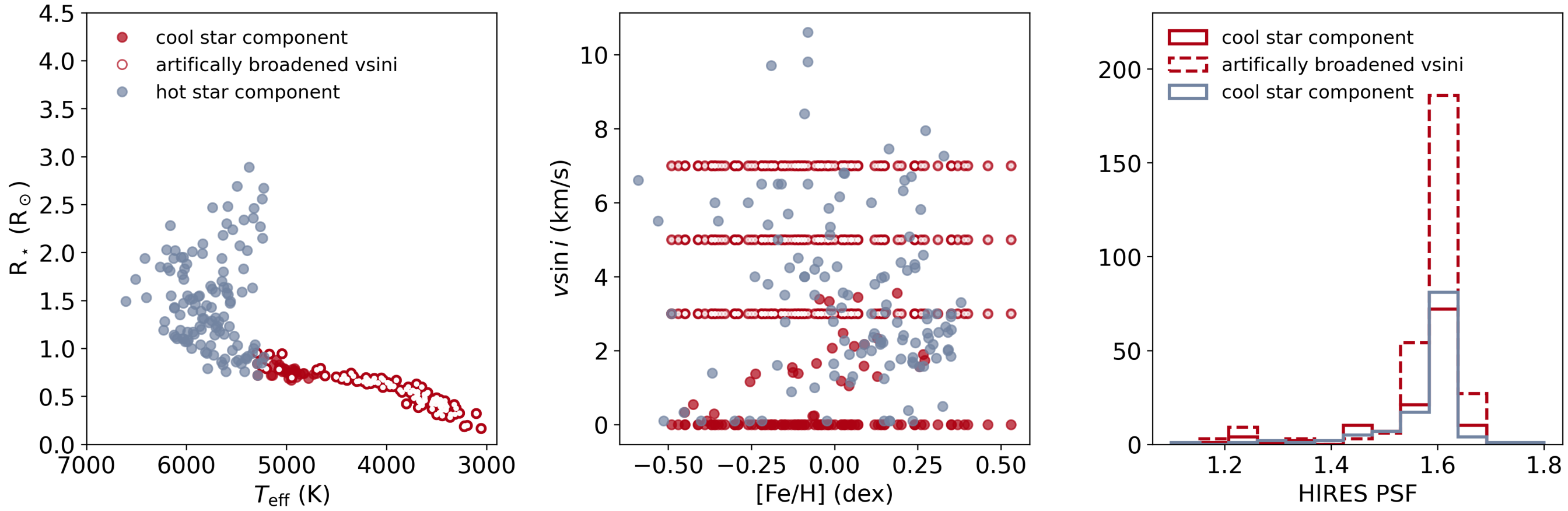

**Figure 1.** Distribution of `Specmatch-Emp` stellar labels—$T_{\rm eff}$, $R_\star$, [Fe/H], and $v \sin i$—as well as PSF values for both of our model training sets. The blue and red points indicate whether the star was part of the hot or cool components of our piecewise Cannon models, respectively. Stars that we included artificially broadened copies of in our cool star Cannon model are filled in with white in the left and middle panels and are shown as a dashed histogram in the right panel. Stars with $R_\star < 2R_\odot$ and $5200 < T_{\rm eff} < 5300$ K are included in both training sets.

**Table 1**
Samples Used in This Analysis

| Sample Name | Source | Purpose | (S/N) | $R$ | Number of Stars | Temperature Range |
|---|---|---|---|---|---|---|
| `SpecMatch-Emp` | S. W. Yee et al. (2017) | Model training set | 150 | ⩾45,000 | 251 | 3056–6608 K |
| CKS | E. A. Petigura et al. (2022) | Survey sample | ⩾20 | ⩾60,000 | 1332 | 3395–6538 K |
| Binary | K. Sullivan et al. (2024) | Binary validation sample | ⩾20 | ⩾60,000 | 97 | 3203–6838 K |
| Single Star | D. Raghavan et al. (2010) | Single-star validation sample | 150 | ⩾45,000 | 102 | 4879–6145 K |

**Note.** The binary sample consists of the subset of stars in the CKS sample that appeared in the binary catalog published in K. Sullivan et al. (2024), as described in Section 4.2. For these stars, we include temperatures of the individual primary and secondary stars in the temperature range.

are publicly available have undergone a preliminary step to remove their instrumental blaze function, as we describe below. We then used the `SpecMatch-Emp` code to shift the spectra into the rest frame and applied a second wavelet-based filtering treatment to remove residual low-frequency variations in the spectra, as detailed in the remainder of this section.

#### *2.2.1. Blaze Function Removal, Part 1: Empirical Models*

The HIRES spectra were processed according to standard practices of the California Planet Search (CPS; A. W. Howard et al. 2010). This procedure removes large-frequency-scale flux variations due to the instrumental blaze function from each individual spectrum, by dividing it by a composite spectrum of "spectrally flat" rapidly rotating B star spectra. However, we noticed that many of the HIRES spectra still contained low-frequency variations, especially for spectra with lower S/Ns. We suspect that this is due to either a mismatched point-spread function (PSF) or polarization between the target spectrum and the composite spectrum. In any case, this necessitated a second processing step, to remove the residual blaze function, which we describe further in Section 2.2.3.

#### *2.2.2. Spectral Registration*

The spectra output by the CPS pipeline are shifted due to the stars' line-of-sight velocities and need to be shifted into the rest frame for the purposes of our analysis. We used `SpecMatch-Emp` (S. W. Yee et al. 2017) to shift the spectra and register them back onto a common wavelength scale. Briefly, the spectra are shifted using a bootstrapping procedure that performs a cross-correlation of the target spectrum with a ladder of five rest-frame-shifted reference spectra. The target spectrum is then shifted according to the reference spectrum that most closely resembles the target spectrum, identified as the spectrum with the largest peak in its cross-correlation function. Next, the spectra are registered onto a common wavelength scale, which we chose to be the standard CKS rest-frame wavelength solution (E. A. Petigura et al. 2017). This produces a spectrum with 16 orders of 4021 pixels each. Finally, in order to produce a 1D spectrum suitable for The Cannon, we clipped each order by 390 pixels on each side, to ensure zero overlap between adjacent orders, to produce a final spectrum of 16 individual orders of 3241 pixels.

#### *2.2.3. Blaze Function Removal, Part 2: Wavelet Filtering*

The processed HIRES observations are not strictly repeatable and often contain nonastrophysical variations on different spectral scales that we wish to remove. To accomplish this, we used a wavelet-filtering process to decompose our signal into variations of different scales, allowing us to experiment with excluding variations that occur on larger scales than the typical spectral line (i.e., continuum-level variations) as well as variations comparable to or smaller than the typical spectral line. Our ultimate goal was to simultaneously minimize the instrumental information and maximize the astrophysical information in our spectra. This method has been used previously to remove systematics from transit signals (M. C. Stumpe et al. 2014; D. del Ser et al. 2018). In practice, it is similar in nature to a Fourier decomposition—the key difference is, instead of decomposing the spectrum into sine and cosine components, which describe oscillations that persist over all wavelengths, wavelet filtering decomposes the spectrum into wavelet components, which describe oscillations that are localized in wavelength space.

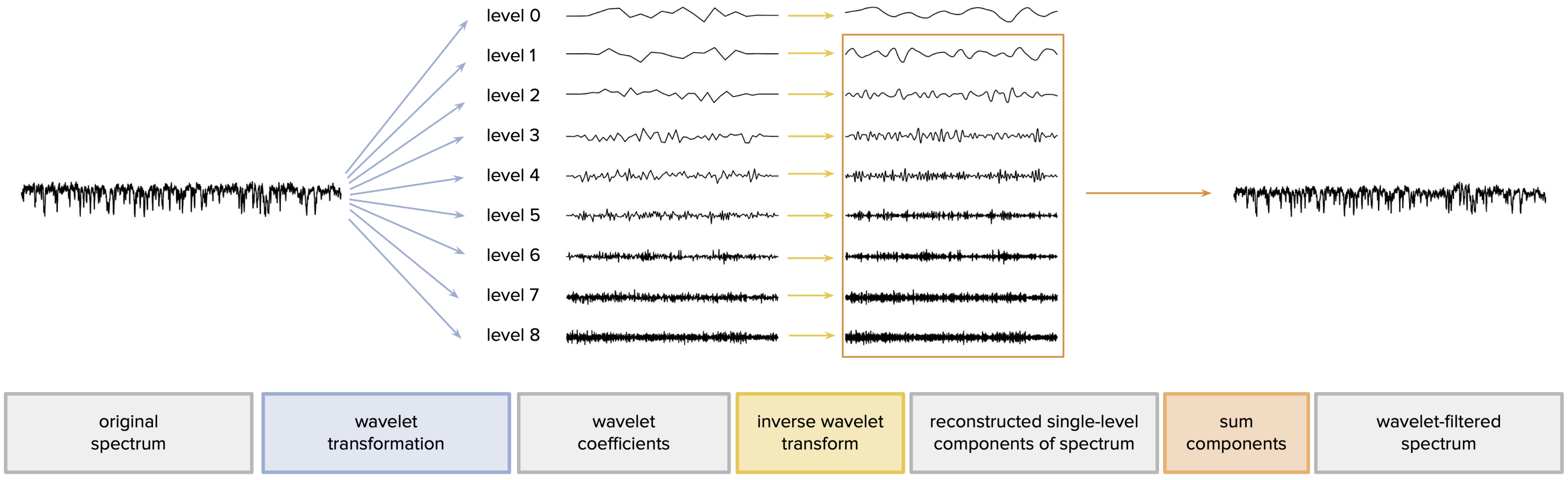

**Figure 2.** Schematic of our wavelet-filtering process to remove low-frequency nonastrophysical variations from our HIRES spectra. The individual HIRES spectrum orders are decomposed into wavelet coefficients via discrete wavelet decomposition. We then reconstruct the signal via wavelet recombination (i.e., inverse wavelet decomposition), excluding the approximation coefficients associated with the lowest frequencies, to produce a wavelet-filtered spectrum.

Figure 2 shows a visual representation of our wavelet-filtering methods. To filter a single spectrum, we first perform a discrete, multilevel decomposition of each individual spectrum order, using the `pywavelets` package. In the same way that a Fourier power spectrum describes the amplitude of a sine wave with a specific frequency, wavelet-transform coefficients describe the amplitude of a wavelet function with a specific frequency at a specific wavelength. Thus, the coefficients of a discrete wavelet transform correspond to the amplitude of a ladder of wavelets with discrete frequencies, evaluated as a function of wavelength. We then recombine the decomposed levels, leaving out the level associated with the lowest-frequency wavelet, where we expect continuum information to be contained. We experimented with different wavelet modes and excluding different combinations of levels in our signal reconstruction. We chose to use a "ym5" wavelet mode and leave out only the lowest-frequency "approximation" coefficients (variations on the order of $\sim$20 Å), because we found that this produced spectra with minimal differences across multiple exposures, thereby preserving only the star's time-invariant astrophysical signal. The output of our wavelet-filtering process is a filtered signal that contains information from all but the lowest-frequency modes of the original spectrum.

We emphasize that unlike traditional continuum treatment procedures, wavelet filtering is not designed to produce flattened spectra normalized to unity. Instead, it produces signals centered at zero, which do exhibit some low-frequency modulations. The key difference between the filtered and unfiltered spectra is that the modulations of the filtered spectra are homogeneous and do not differ from star to star and thus do not provide any information about the star's physical properties or other systematics that might compromise the information our model learns during its training step. This is demonstrated in Figure 3, where we show two spectra of KOI-99 taken on different nights, where any differences should come from instrumental systematics. We plot the ratio of the spectra before and after wavelet filtering, as well as the ratio of the filtered-out wavelet components. As we can see from this figure, the ratio of the wavelet-filtered spectra is smaller than the ratio of the unfiltered spectra, indicating that the wavelet-filtered spectra vary less on a night-to-night basis compared to their unfiltered counterparts.

We note that it is possible for some spectral information to be lost in this process, particularly at wider, deeper features that may be captured by the level 0 coefficients that we remove during the filtering process. Indeed, for the example spectrum in Figure 3, we see that the relative depths of the three deepest lines are slightly altered during this process. As long as the line depths retain the necessary astrophysical information and remain consistent across stars of similar type, this transformation does not necessarily inhibit the ability to model a realistic spectrum when comparing wavelet-filtered models to wavelet-filtered data. The success of the label transfer suggests this is the case. Nevertheless, we caution against applying these methods beyond label transfer without further analysis to ensure the viability of wavelet-based filtering in a broader context. While a more in-depth exploration of the applications of wavelet-filtered spectra is beyond the scope of this paper, we discuss the potential impacts of this on our results in Section 5.

We proceed with our analysis by using both spectra that are output by `SpecMatch-Emp` (hereafter, the "unfiltered" spectra) and spectra that have gone through an extra step of wavelet filtering and by comparing our model performance for both cases.

#### 2.2.4. Pixel Mask

The HIRES spectra are contaminated by telluric absorption features imprinted by molecules in the Earth's atmosphere. These features come from outside the target star reference frame; as a result, they are shifted out of the rest frame when the rest of the spectrum is shifted into the rest frame and manifest at different pixels in the HIRES spectrum. We masked pixels prone to telluric contamination, both to assure the quality of our training set spectra and to avoid the model overfitting to telluric lines at the cost of producing a good fit elsewhere during its test step (see Section 2.3). We referenced a list of telluric lines,[3] defined as the minimum and maximum wavelengths of each telluric feature presented in the HIRES *r*- chip bandpass. We added a $30\,\mathrm{km\,s^{-1}}$ padding to each wavelength range, to account for the shifting of telluric lines due to barycentric corrections, and we added another $100\,\mathrm{km\,s^{-1}}$ padding, to account for the shifting due to the relative radial velocities of the stars before rest-frame shifting. Our final telluric mask is illustrated in Figure 4. Because The Cannon fits to each pixel individually in its training step, we do not mask telluric lines when training the model. Instead, we mask telluric lines during the test step, by fitting to only the unmasked pixels in the spectrum, as described further in Section 2.3.

[3] Available at https://www2.keck.hawaii.edu/inst/common/makeewww/Atmosphere/.

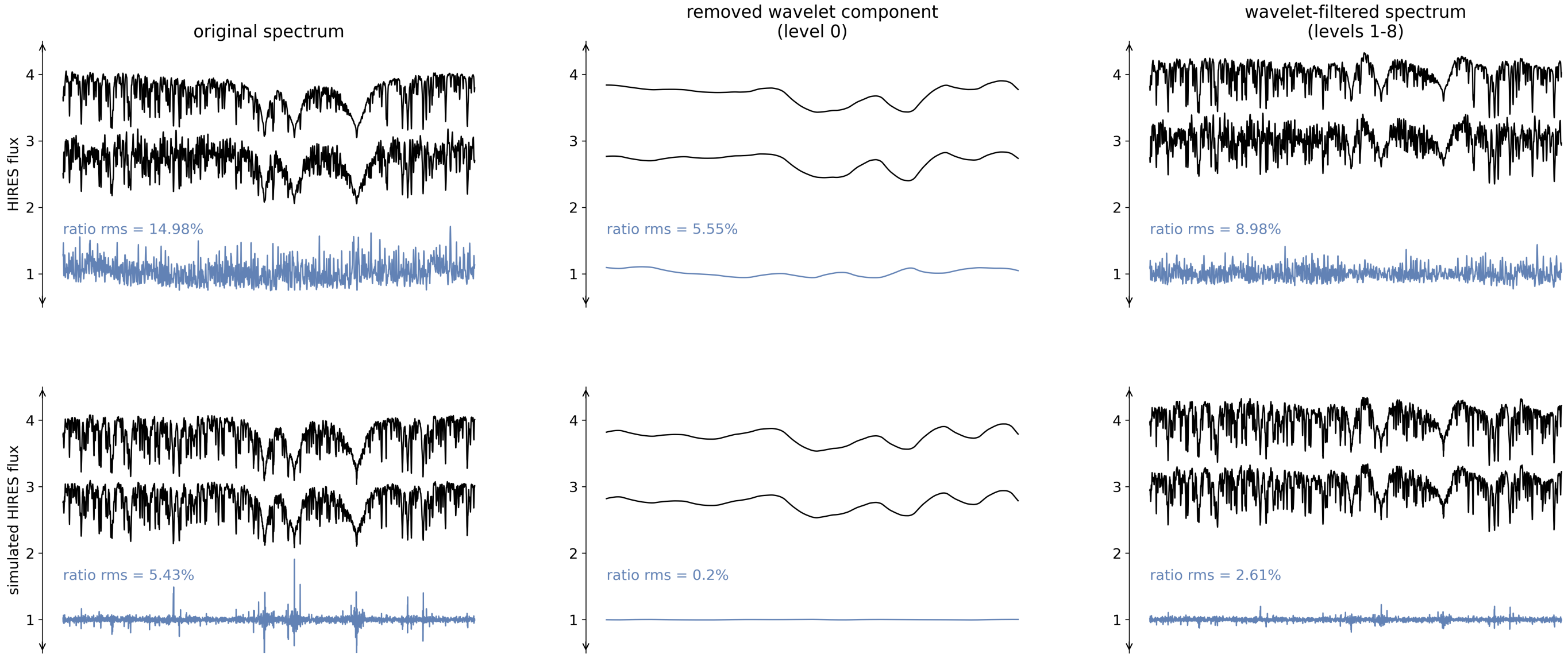

**Figure 3.** Demonstration of the ability of our wavelet-filtering process to remove nonastrophysical night-to-night variations from HIRES spectra. The top panel shows snippets spanning 5130–5200 Å for two HIRES spectra of KOI-99 taken on different nights, where variations between the two spectra should be primarily due to instrumental effects. The bottom panel shows two simulated KOI-99-like spectra, where variations between the two spectra are due to injected noise only. While the wavelet filtering does not produce flat spectra, like traditional continuum treatments, it is effective at removing low-frequency night-to-night variations present in the HIRES spectra while simultaneously preserving astrophysical information in both real and simulated spectra.

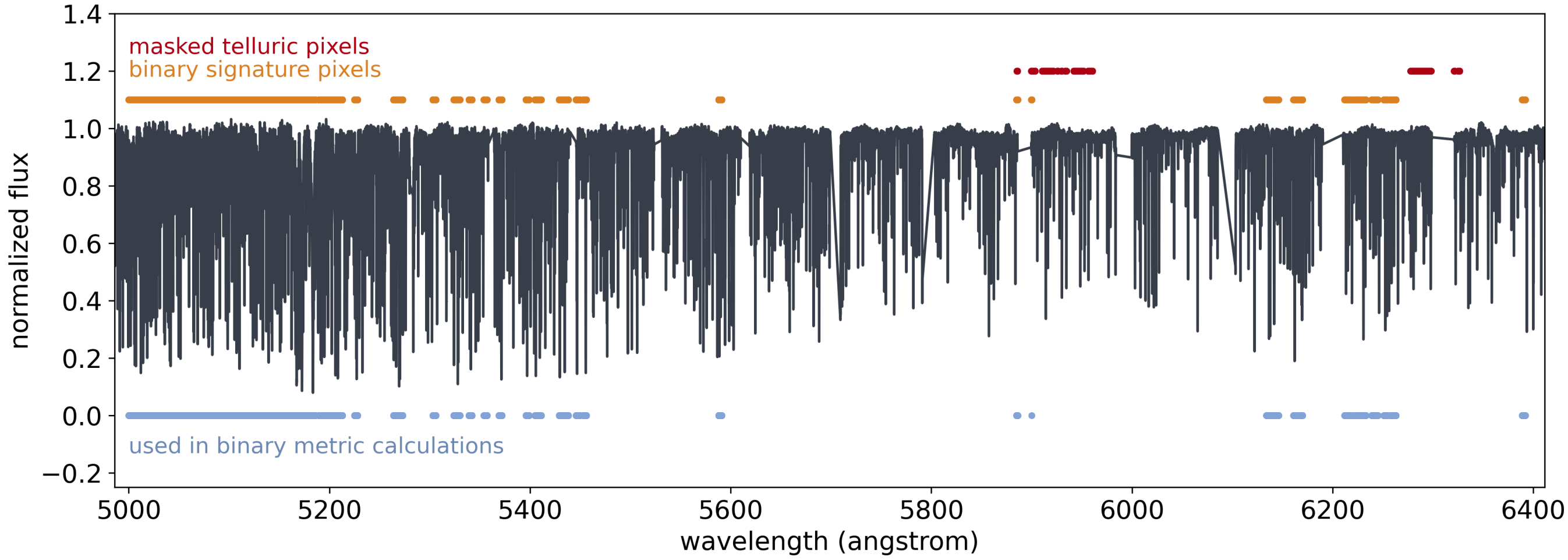

**Figure 4.** HIRES spectrum of sample star HD 106088. The pixels we mask due to telluric contamination are shown in red, and the binary signature pixels where we expect the strongest signatures of a secondary star are shown in orange. The binary signature pixels that are not contaminated by tellurics—shown in blue at the bottom—are the pixels we used to calculate the ΔBIC associated with binary detection (Section 4).

We expect our model to be more sensitive at specific features where the signature of the secondary is more prevalent. Along these lines, wavelengths where the single-star and binary spectra are expected to be the same effectively add noise to any goodness-of-fit comparison between the two models. Thus, we incorporated a second "binary signature" pixel mask, using the methodology described below, which includes only pixels expected to show the strongest signatures of the secondary star.

To determine which pixels to include in our binary signature pixel mask, we used ab initio models to simulate a coarse grid of 20 binary spectra with $T_{\mathrm{eff1}} = 4000$–$7000$ K and $T_{\mathrm{eff2}} = 3000 - T_{\mathrm{eff1}}$, following the methods outlined in Section 5.1. We calculated the difference between the binary spectrum and the primary spectrum for each binary to generate a grid of difference spectra. We then selected pixels from this grid whose average absolute difference or standard deviation were in the top 20th percentile of all pixels—this corresponds to an average difference and standard deviation of 0.004 and 0.01 in normalized flux units, respectively. These pixels (hereafter, “binary signature pixels”) represent the HIRES spectrum pixels that exhibit either the largest difference between the binary and single-star spectra or the largest scatter in difference between the binary and single-star spectra. This binary signature pixel mask will be used in the model comparison by computing the goodness-of-fit metrics of the best-fit single- and binary star models using only the selected pixels, as discussed further in Section 4.3.

### 2.3. Single-star Model

We used a data-driven algorithm called The Cannon (M. Ness et al. 2015) to train a generative model that predicts the per-pixel flux of single stars in HIRES as a function of labels $l_n = (T_{\rm eff}, R_\star, [{\rm Fe/H}], v \sin i$, and PSF), where the PSF is the instrumental PSF associated with each observation. While other models of HIRES spectra have included more training labels, it has been shown that $T_{\rm eff}$, $R_\star$, [Fe/H], and $v \sin i$ are sufficient for the purposes of binary detection (K. El-Badry et al. 2018a, 2018b). We included the instrumental PSF as an additional training label, to correct for differences in the PSF across different observations in our training set, which compromise our models’ ability to accurately infer $v \sin i$ when left unaccounted for. Holding all other stellar labels constant, decreasing the resolution of a spectrum broadens the spectral lines, which is approximately similar to increasing $v \sin i$ for slowly rotating stars. Training our model on both $v \sin i$ and the observation-specific PSF allows it to learn how the flux varies with both $v \sin i$ and the PSF, to produce a spectrum that is more accurately tethered to the star’s physical $v \sin i$ value.

We used the implementation of The Cannon developed by A. R. Casey et al. (2016)[4] to train our spectral model. A detailed description of these methods can be found in M. Ness et al. (2015) and A. R. Casey et al. (2016), but we provide a brief review here. In short, The Cannon assumes that the flux of a star $n$ at pixel $j$ and the stellar labels $l_n$ obey the following relationship:

$$f_{nj} = \boldsymbol{v}(l_n) \cdot \theta_j + e_{nj}, \tag{1}$$

where $\boldsymbol{v}(l_n)$ is a “vectorizing” function that describes how the flux is assumed to vary with the labels, and $e_{nj}$ are the flux errors, which are sampled from a Gaussian with zero mean and variance equal to the instrumental uncertainty and intrinsic pixel scatter of the model summed in quadrature, $\sigma_{nj}^2 + s_j^2$. For our model, we assume the relationship between the pixel flux and stellar labels is a second-order linear relationship with polynomial coefficient vector $\theta_j$, where $\theta_j$ is a $21 \times 1$ matrix of label coefficients and their cross-terms ($T_{\rm eff}$, $R_\star$, $T_{\rm eff}^2$, $R_\star^2$, $T_{\rm eff}R_\star$, etc.). In other words, we model the flux as a polynomial function of the stellar labels, plus some deviation sampled from a Gaussian with zero mean and variance equal to the observation and model uncertainty summed in quadrature.

The Cannon uses training spectra during its “training step” to develop a spectral model, and it then uses this model to fit to individual spectra outside the training set during its “test step.” The training step of The Cannon determines the polynomial coefficients and intrinsic model scatter $(\theta_j, s_j^2)$ that minimize the following function:

$$\theta_j, s_j^2 = \arg\min_{\theta_j, s_j^2} \sum_{n=1}^{N_{\rm stars}} - \ln p\,(y_{nj}|\theta_j, l_n, s_j^2), \tag{2}$$

where $y_{nj}$ is the observed flux of the training set star $n$ at pixel $j$, and $\ln p$ describes the per-pixel log-likelihood of the model flux:

$$\ln p\,(y_{nj}|\theta_j, l_n, s_j^2) = -\frac{[y_{nj} - \boldsymbol{v}(l_n) \cdot \theta_j]^2}{s_j^2 + \sigma_{nj}^2} - \ln\,(s_j^2 + \sigma_{nj}^2). \tag{3}$$

The final output of the training step is a set of polynomial coefficients $\theta_j$ and per-pixel intrinsic model scatter $s_j^2$ for each spectrum pixel that best describe how the flux of the training set stars depends on the stellar labels.

Once the model is trained and the coefficients are determined, we can use the model in the test step to estimate the labels of spectra outside the training step. This step typically varies only the stellar labels in $l_n$ to determine the the best-fit Cannon model. However, we noticed during our analysis that many of the HIRES spectra in our samples show large residuals between the spectra and their best-fit single-star models, due to small errors in rest-frame shifting carried out by `SpecMatch-Emp` (see Section 2.2.2). Thus, we subsequently apply a relative radial velocity $v_n$ to The Cannon model output to predict $g_{nj'}$—i.e., the flux corrected for imperfect rest-frame shifting during data processing:

$$g_{nj'}(l_{{\rm single},n}) = \boldsymbol{D}(j', j, v_n) \cdot [v(l_n) \cdot \theta_j]. \tag{4}$$

Here, $\boldsymbol{D}(j', j, v_n)$ is an operator that applies a Doppler shift by the relative velocity $v_n$ and interpolates The Cannon rest-frame flux $v(l_n) \cdot \theta_j$ from the $j$ Cannon model grid to the $j'$ HIRES pixel grid, and $l_{{\rm single},n} = (l_n, v_n)$ describes the full set of six single-model parameters that describe a particular star $n$. We note that while these shifting errors are present in the training data as well, we experimented with training a second model on the velocity-corrected spectra and found that it does not substantially affect our model performance.

We estimate the best-fit model for a given spectrum by determining the model parameters in $l_{{\rm single},n}$ that maximize the model likelihood function (or, rather, minimize the negative likelihood function), evaluated at all nontelluric pixels (see Section 2.2.4 for a description of our telluric mask):

$$\hat{l}_{{\rm single},n} = \arg\min_{l_{{\rm single},n}} - \ln \mathcal{L}\,(y_n|l_{{\rm single},n}), \tag{5}$$

where $\ln \mathcal{L}\,(y_{nj'}|\theta_{j'}, l_{{\rm single},n}, s_{j'}^2)$ (hereafter, $\ln\mathcal{L}$) represents the Bayesian likelihood associated with a particular set of parameters $l_{{\rm single},n}$ for a HIRES pixel $j'$:[5]

$$-\ln \mathcal{L} = \frac{1}{2} \sum_{j'=1}^{N_{\rm pixels}} \frac{[y_{nj'} - g_{nj'}(l_{{\rm single},n})]^2}{s_{j'}^2 + \sigma_{nj'}^2} + \log 2\pi (s_{j'}^2 + \sigma_{nj'}^2). \tag{6}$$

[4] Available at https://github.com/andycasey/AnniesLasso.

[5] For our optimization, we approximate $\theta_{nj'}$ and $s_{j'}^2$ as $\theta_{nj}$ and $s_j^2$, respectively. We recognize that this approximation is not technically correct; however, we find that this is a workable approximation in the low-Doppler-shift regime we are modeling.

To do this, we first conduct a brute search across the H-R diagram, fixing all labels except $T_{\rm eff}$ and $R_\star$ to determine the ballpark values of these labels for the star of interest. We then initialize an optimizer at these values and vary all six model parameters using the Nelder–Mead method in `scipy.optimize.minimize()`. We constrain the optimizer to only sample labels that are within the bounds of the training set (Figure 1), allowing $v_n$ to vary from −20 to 20 km s$^{-1}$. We also reparameterize $v \sin i$ to $\log(v \sin i)$ to avoid fitting $v \sin i < 0$.

The Cannon's quadratic approximation of $f_{nj}$ has been shown to be sufficient for training sets spanning a $T_{\rm eff}$ range of ∼2000 K; however, it has been suggested that broader applications may require increased model complexity (M. Ness et al. 2015). Since our training set spans a larger range of parameter space ($T_{\rm eff}$ = 3000–7000 K), we split our training set into two separate training sets of hot ($T_{\rm eff}$ > 5200 K) and cool ($T_{\rm eff}$ < 5300 K) stars, and we trained two models that span smaller temperature ranges over which we expect the quadratic approximation to be more accurate. We allowed for 100 K overlap for the two components but removed stars with $R_\star > 2$ in the overlapping region from the cool star sample to avoid a training set with large gaps in the training set parameter space (see Figure 1). To carry out The Cannon's test step, we fit each spectrum to both the cool- and hot-star models and select the fit with the larger $\ln\mathcal{L}$ (Equation (6)), effectively fitting a piecewise Cannon model to each spectrum.

We note that another way to achieve increased model complexity is by training a neural network that assumes a more flexible relationship between the stellar flux and labels of the training set (e.g., The Payne—Y.-S. Ting et al. 2019; The Data-Driven Payne—M. Xiang et al. 2019). We experimented with using The Payne to train a neural network but found that for our training set, the models were prone to overfitting to training set spectra at the cost of providing a poor fit to spectra outside the training set. We expect that this poor performance is likely due to the small number of spectra in our training set. A more in-depth exploration of The Payne and its applications to HIRES spectra is beyond the scope of this paper.

### 2.4. Binary Model

We used our piecewise Cannon model to construct a composite binary model that predicts the flux of spectral binaries as a function of the labels associated with the primary and secondary stars. We model the spectral binary flux as a linear combination of the Doppler-corrected primary and secondary star flux:

$$g_{{\rm binary},j'} = W_1 \times g_{1j'}(l_{\rm single,1}) + W_2 \times g_{2j'}(l_{\rm single,2}), \tag{7}$$

where $l_{\rm single,1}$ and $l_{\rm single,2}$ are the respective label vectors of the primary and secondary star, and $W_1$ and $W_2$ are flux-weighted prefactors calculated based on the flux ratio of the primary and secondary stars:

$$W_1 = \frac{f_1}{f_1 + f_2} = \frac{f_{\rm rel}}{f_{\rm rel} + 1}, \tag{8}$$

$$W_2 = 1 - W_1. \tag{9}$$

Here, $f_{\rm rel}$ is the flux ratio of the primary and secondary stars, $f_{\rm rel} = f_1/f_2 = 10^{(m_2 - m_1)/2.5}$, and $m_1$ and $m_2$ are their respective $V$-band magnitudes. We estimated $m_1$ and $m_2$ based on the empirical $T_{\rm eff}$–magnitude relations for main-sequence stars published in M. J. Pecaut & E. E. Mamajek (2013). We note that these empirically derived flux ratios are approximated as single values across the full HIRES wavelength range. In reality, the flux ratio between the primary and secondary stars is wavelength-dependent. We experimented with using blackbody models to calculate the binary flux ratios and found that the variations in the flux ratios across the full spectrum were anywhere from 1% to 3%, comparable to the ≲3% uncertainties in the blackbody flux ratios that arise from uncertainties in stellar radii. These uncertainties are smaller on average than the ∼2%–5% accuracy of our single-star model (Section 4.3); thus, we do not expect our approximation of a constant flux ratio to significantly affect our analysis.

While we are primarily interested in identifying spectral binaries with <10 km s$^{-1}$ velocity offsets, we allow for the individual velocities of the primary and secondary ($v_1$ and $v_2$) to float, to account for both imperfect rest-frame shifting of the spectrum and 1–10 km s$^{-1}$ velocity offsets between the primary and secondary that would have been missed by previous searches for SB2 signatures in CKS spectra (i.e., R. Kolbl et al. 2015).

We vary the following labels and velocities of the primary and secondary stars:

$$\begin{aligned} l_{\rm binary} = (&T_{\rm eff1}, R_{\star 1}, [{\rm Fe/H}]_1, v\sin i_1, v_1, {\rm PSF}_1, \\ &T_{\rm eff2}, R_{\star 2}, v\sin i_2, v_2); \end{aligned} \tag{10}$$

and we assume that the metallicity [Fe/H] and the instrumental PSF are the same for both the primary and secondary star to derive $l_{\rm single,1}$ and $l_{\rm single,2}$ and compute our binary model prediction $g_{{\rm binary},j'}$. We determine the parameters $l_{{\rm binary},n}$ associated with the best-fit binary model for a particular star $n$ using a similar optimization procedure as for our single-star optimization:

$$\hat{l}_{{\rm binary},n} = \underset{l_{\rm binary}}{\arg\min} - \ln\mathcal{L}_{\rm binary}\,(y_n | l_{{\rm binary},n}), \tag{11}$$

where $\ln\mathcal{L}_{\rm bin}$ represents the Bayesian log-likelihood associated with a particular set of binary model parameters:

$$\begin{aligned} -\ln\mathcal{L}_{\rm binary} = \frac{1}{2}\sum_{j'=1}^{N_{\rm pixels}} &\frac{[y_{nj'} - g_{{\rm binary},j'}(l_{{\rm binary},n})]^2}{s_{{\rm binary},j'} + \sigma_{nj'}^2} \\ &+ \log 2\pi(s_{{\rm binary},j'} + \sigma_{nj'}^2), \end{aligned} \tag{12}$$

and $s_{{\rm binary},j'}$ is the intrinsic scatter of the binary model, which we estimate as a linear combination of the intrinsic scatter of the piecewise Cannon components used to model the primary and secondary stars ($s_{1j'}^2$ and $s_{2j'}^2$, respectively), weighted according to their relative flux contributions:

$$s_{{\rm binary},j'} = W_1 \cdot s_{1j'}^2 + W_2 \cdot s_{2j'}^2. \tag{13}$$

We make two further modifications to our optimization procedure (Section 2.3) to adapt it for the binary scenario. First, we perform the optimization in three separate regimes of our piecewise Cannon model: (1) the primary and secondary are modeled by the hot piecewise component; (2) the primary and secondary are modeled by the cool piecewise component; and (3) the primary is modeled by the hot piecewise component and the secondary is modeled by the cool piecewise component. As with the single-star optimization, we select the best-fit binary model as the model associated with the highest $\ln\mathcal{L}$ across all three regimes.

The second modification we make is to the initial brute optimization. To determine the best-fit binary in a particular regime of our piecewise Cannon model, we first conduct a brute search varying only $T_{\rm eff1}$ and $T_{\rm eff2}$ within the bounds of their respective piecewise component training sets. We then initialize an optimizer at these values and vary all 10 model parameters ($T_{\rm eff}$, $R_\star$, [Fe/H], vsin$i$, the PSF of the primary, $T_{\rm eff}$, $R_\star$, the $v \sin i$ of the secondary, $v_1$, and $v_2$) using the Nelder–Mead method in `scipy.optimize.minimize()`. Similar to the single-star optimization, the labels of the primary and secondary are constrained to within the bounds of their respective piecewise Cannon model components.

## 3. Validation of Label Transfer

Historically, label transfer has been a prominent application of data-driven spectroscopic models (e.g., M. Ness et al. 2015; A. Y. Q. Ho et al. 2017; S. W. Yee et al. 2017; Y.-S. Ting et al. 2019; M. Rice & J. M. Brewer 2020). Briefly, label transfer leverages the fact that for a given survey, a subsample of stars typically have derived labels with higher fidelity from a previous survey. The process of label transfer involves first learning the relationship between the flux and "ground-truth" stellar labels for the subset of well-characterized stars (i.e., the model training set) and then using this model to infer stellar labels for a larger population.

To assess our model's ability to infer stellar labels, we performed a leave-20%-out validation of our piecewise Cannon model. For each 20% subset of the full training set, we trained a new model on the remaining 80% and used this model to compute labels of the held-out subset to obtain Cannon-inferred labels for all of the stars in our training set. We do this for two models: one trained on unfiltered training spectra and one trained on wavelet-filtered training spectra (see Section 2.2.3). The left column of Figure 5 shows The Cannon output labels plotted versus the `SpecMatch-Emp`-reported labels, which we treat as the ground truth. The rms scatter around the one-to-one line represents our estimated $1\sigma$ label uncertainties for our model. As we can see from the figure, wavelet filtering improves our model's ability to predict labels that agree with the ground-truth labels. The model trained on unfiltered spectra agrees with ground-truth labels to within 84 K in $T_{\rm eff}$, 0.18 $R_\odot$ in $R_\star$, 0.11 dex in [Fe/H], 1.01 km s$^{-1}$ in $v \sin i$, and 0.07 in the instrumental PSF. On the other hand, the model trained on the wavelet-filtered spectra shows agreements of 27 K in $T_{\rm eff}$, 0.10 $R_\odot$ in $R_\star$, 0.09 dex in [Fe/H], 0.87 km s$^{-1}$ in $v \sin i$, and 0.05 in the instrumental PSF with the ground-truth spectra. These precisions are comparable to the model precisions of existing data-driven models of HIRES spectra (S. W. Yee et al. 2017; M. Rice & J. M. Brewer 2020) and demonstrate the significant improvement in model performance when we apply wavelet filtering. We also see this improvement reflected in the Pearson $r$ coefficient of the labels plotted in Figure 5—for all labels, the $r$ coefficient of the predicted and ground-truth labels is larger and indicative of a stronger, more positive correlation.

The improvement from wavelet filtering is even more significant for spectra with more moderate S/N. We used our piecewise Cannon models to compute labels of the S/N $\sim$ 45 stars in our CKS survey sample (described in further detail in Section 4.1). The right column of Figure 5 shows the unfiltered Cannon output labels inferred from unfiltered CKS spectra plotted versus the "ground-truth" CKS labels reported in E. A. Petigura et al. (2022) in yellow. The wavelet-filtered Cannon output labels inferred from wavelet-filtered CKS spectra are plotted in blue.

We see from Figure 5 that the improved agreement with the ground-truth labels is even more stark for these spectra than for the training spectra. We also see that the unfiltered Cannon model produces unreliable [Fe/H] labels for a number of CKS spectra, whereas the filtered model does not have this issue. We conclude from these results that our wavelet-filtering process has the potential to drastically improve the quality of the labels derived from data-driven models for spectra with S/N $\sim$ 45. This is likely due to the shorter average exposure times for low-S/N observations, which are correlated with larger blaze function variability that is more difficult to remove with the preliminary removal methods detailed in Section 2.2.1.

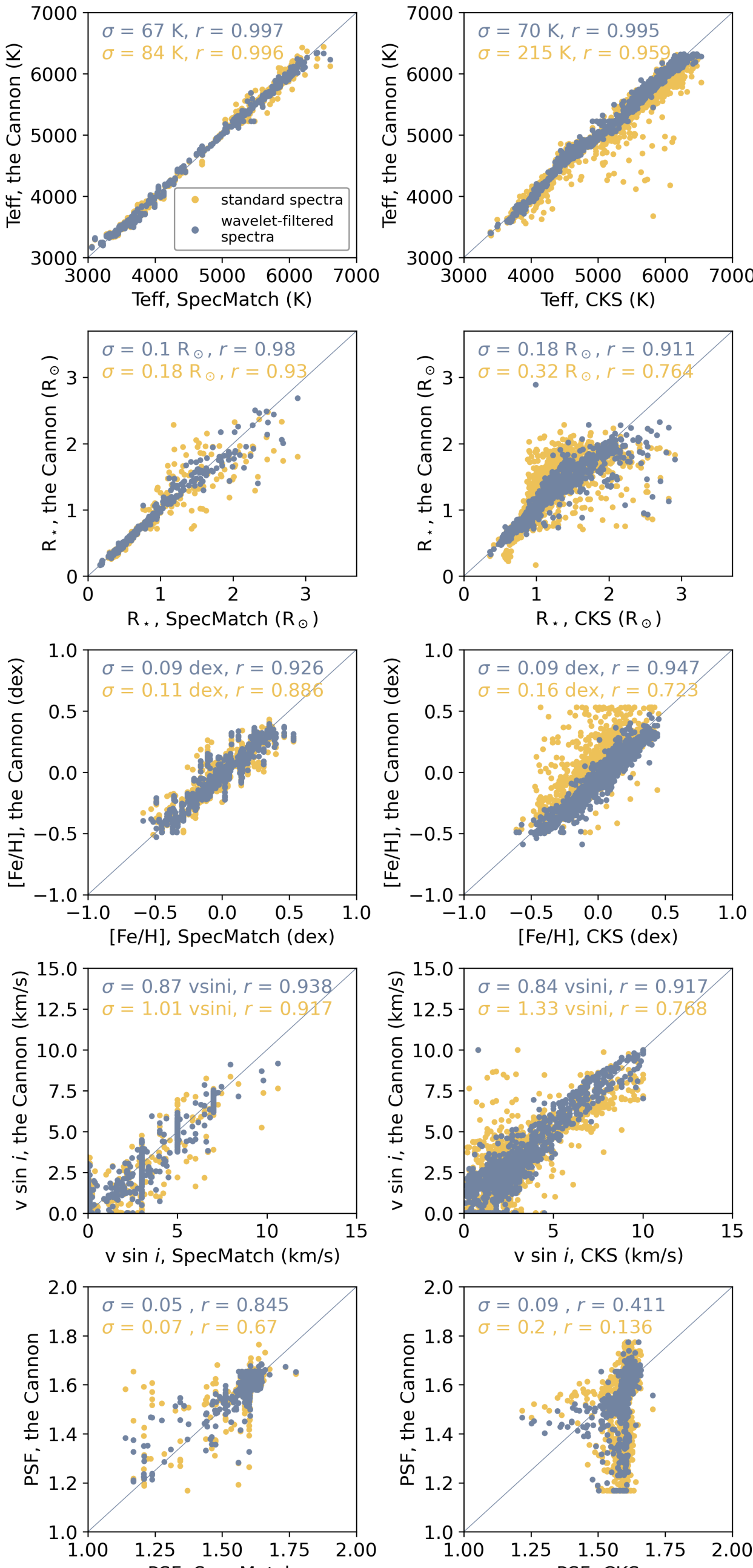

**Figure 5.** Left: labels of our `SpecMatch-Emp` training set spectra (S/N $\approx$ 150) were computed by our Cannon model, using leave-20%-out validation for both the original and wavelet-filtered models. The corresponding rms scatter and Pearson $r$ coefficient are shown for both cases. Right: demonstration of model label transfer performance on moderate-S/N CKS spectra (S/N $\gtrsim$ 45). The labels of our CKS samples were computed by our Cannon models for CKS spectra using both unfiltered (yellow) and wavelet-filtered (blue) spectra. Wavelet filtering significantly improves our Cannon model agreement with the ground-truth labels in both cases, but the improvement is more drastic at moderate S/N.

## 4. Sensitivity to Binaries

We have demonstrated that our model is effective at determining the stellar labels of stars based on their HIRES spectra. In this section, we explore our model's ability to detect signatures of unresolved binaries in these same spectra. As described below, this application of data-driven models is more demanding and requires more accurate per-pixel flux estimations of stellar spectra than tested and proven label transfer applications.

### 4.1. Survey Data

Our ultimate goal is to use our data-driven spectral binary emulator to distinguish between HIRES spectra of single stars and spectral binaries from the CKS sample of known planet hosts. This sample includes HIRES spectra of 1305 planet hosts with surface temperatures of 4700–6500 K (CKS DR1; E. A. Petigura et al. 2017), as well as 411 spectra of cooler planet hosts with $T_{\rm eff}$ = 3500–5000 K (CKS DR2; E. A. Petigura et al. 2022), for a total of 1712 stars. These spectra have a typical per-pixel S/N of 45 and spectral resolution of $R \geqslant 60{,}000$; however, the exact S/Ns and spectral resolutions vary for different objects.

Previous studies have identified unresolved binaries in this sample—that is, targets with companions with separations within the $0\farcs8$ HIRES slit width. In particular, 22 stars were flagged as spectroscopic binaries in E. A. Petigura et al. (2022). Another 132 were listed as binaries identified by Kepler follow-up imaging programs and further characterized in K. Sullivan et al. (2024), as detailed in the following section. Of the remaining targets, 59 were reported in a KOI binary catalog that combined results from Keck II/NIRC2 imaging, UKIRT Hemisphere Survey photometry, and a patchwork of existing literature. This catalog will be described further in A. L. Kraus et al. (2025, in preparation), and many targets have already been published in A. L. Kraus et al. (2016). We removed these binaries from our survey sample, along with 143 stars with reported labels outside our model training set range, and we reserved the remaining 1332 stars for our analysis, including our label transfer performance assessment in Section 3.

### 4.2. Validation Data

We reserved the aforementioned binaries in our CKS sample that appeared in the binary catalog published in K. Sullivan et al. (2024) for the purposes of validating our model's ability to identify spectral binaries. This catalog was assembled from a combination of KOI follow-up imaging programs from E. Furlan et al. (2017) and C. Ziegler et al. (2018) as well as supplemental adaptive optics imaging. The targets were then observed with the LRS2 instrument at McDonald to obtain revised star properties ($T_{\rm eff}$, $R_\star$) for both the primary and secondary star. Of the 132 CKS targets that appeared in the complete binary catalog, 97 have companions that fall within the HIRES slit (angular separation $\leqslant 0\farcs8$), with 74 companions within $0\farcs4$, where we had conservatively expected 100% of the secondary flux to fall within the slit. We compiled all binaries with $\leqslant 0\farcs8$ separations to assemble our binary validation set.

We also composed a sample of single stars to validate our binary model. We selected this sample from a survey carried out in D. Raghavan et al. (2010). This study surveyed a sample of nearby stars with multiple complementary techniques that yielded near 100% completeness to stellar binaries at all separations. To construct our single-star sample, we queried the 173 stars in this survey where no companion was detected, with 153 single stars having available HIRES spectra. We then removed stars with stellar labels outside the bounds of our training set model, based on previously derived stellar properties from the Spectroscopic Properties of Cool Stars catalog (J. M. Brewer et al. 2016). Our final validation sample consists of 102 single stars.

The D. Raghavan et al. (2010) survey targeted stars within 25 pc and is, by construction, much brighter than the Kepler field stars in our binary sample. Our single-star and binary samples have large S/N discrepancies as a result, with average per-pixel S/Ns of 150 and 45 in each respective sample. Despite our incorporation of flux errors into our goodness-of-fit estimation (as described in the next section), we found that the S/N was the best predictor of our model performance across both the single-star and binary validation samples. Thus, in order to obtain a validation sample lending itself to a one-to-one comparison, we degraded each single-star-spectrum S/N by adding Gaussian noise with standard deviation $\sigma_{\rm added} = \sqrt{\sigma^2_{\rm binary} - \sigma^2_{\rm single}}$, where $\sigma_{\rm binary}$ are the flux errors of a randomly selected star from our binary validation sample. Our final validation sample consists of 102 single stars and 97 unresolved binaries with per-pixel S/N $\sim$ 45.

### 4.3. Performance on Validation Data

To assess our model's ability to distinguish between single stars and binaries, we investigated the extent to which the stars in our validation samples are significantly better fit by our binary model than our single-star model. Since the binary model has more free parameters, we expect the $\chi^2$ of the best-fit binary model to be consistently lower than that of the best-fit single-star model. To account for this, we evaluated the goodness of fit of a particular single-star or binary model using the Bayesian information criterion (BIC; see G. Schwarz 1978), which penalizes models with more free parameters:

$$\mathrm{BIC} = k \log N_{\rm pixels} - 2 \log \mathcal{L}, \tag{14}$$

where $k$ is the number of free parameters, $N_{\rm pixels}$ is the number of pixels in the spectrum, and $\ln\mathcal{L}$ is the Bayesian likelihood associated with the best-fit model (Equation (3)). We then calculated the difference between the BICs of the best-fit single-star binary models. A value of $\Delta\mathrm{BIC} > 10$ typically corresponds to a Bayesian evidence ratio of 100 and has been widely adopted as a criterion for strong evidence in favor of the lower-BIC model (see G. Schwarz 1978; A. R. Liddle 2007, and references therein). However, this assumption relies strongly on assumptions that our models do not follow, and a ΔBIC threshold of 10 is less physically meaningful in our case. Thus, we aim to determine a critical ΔBIC value for binary detections empirically based on ΔBIC distributions of known single stars and binaries.

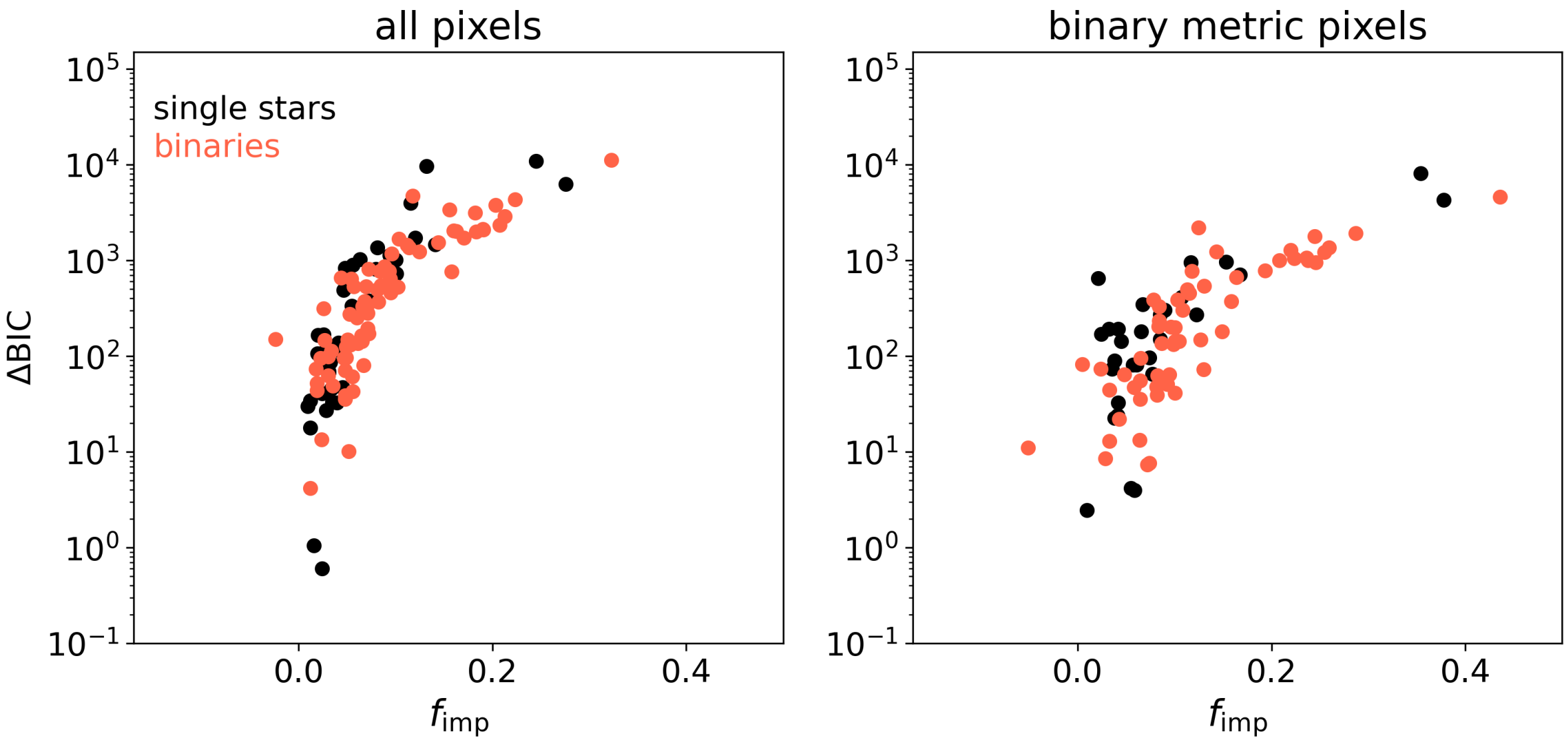

**Figure 6.** ΔBIC plotted versus $f_{imp}$ for all stars in our single-star and binary validation samples. The left panel shows the quantities evaluated at all nontelluric HIRES spectrum pixels, and the right panel shows the quantities evaluated only at our binary signature pixels, where we expect the signature of the secondary star to be the strongest (see Section 5.1). In both cases, our ability to meaningfully distinguish between the single-star and binary samples is limited.

Following the formalism outlined in K. El-Badry et al. (2018b), we also calculated the improvement fraction $f_{imp}$, which measures the degree to which the spectrum is better fit by the binary model, relative to how much it differs from the single-star model:

$$f_{\rm imp} = \frac{\sum[(|g_{{\rm single},j'} - y_{j'}| - |g_{{\rm binary},j'} - y_{j'}|)/\sigma_{j'}]}{\sum[|g_{{\rm single},j'} - g_{{\rm binary},j'}|/\sigma_{j'}]}, \qquad (15)$$

where $y_{j'}$ and $\sigma_{j'}$ are the observed HIRES flux and flux errors, $g_{{\rm single},j'}$ is the best-fit single-star model, and $g_{{\rm binary},j'}$ is the best-fit binary model. We expect the binaries in our validation sample to have both larger ΔBIC and $f_{imp}$ values, indicating significant improvement from the binary model.

Figure 6 shows the ΔBIC and $f_{imp}$ values for stars in our single-star and binary validation samples. We report values that are calculated using all nontelluric pixels, as well as values that are calculated using only pixels in our binary pixel mask. As we can see from this figure, the main advantage of using our binary signature pixel mask is that it reduces the spread of ΔBIC values of our single-star spectra, concentrating most single stars to $10 < \Delta{\rm BIC} < 10^3$. This is in line with what we expect—the mask removes pixels with low information content, effectively removing noise from our analysis. We also see that the pixel mask slightly increases the $f_{imp}$ values of the binaries relative to the single stars.

To further quantify the degree of similarity between our samples, we calculated the two-sided Kolmogorov–Smirnov (K-S) test statistic for the ΔBIC and $f_{imp}$ metrics plotted in Figure 6. The two populations have K-S $p$-values of $3.4 \times 10^{-6}$ and $1.2 \times 10^{-7}$ for the respective ΔBIC and $f_{imp}$ distributions using binary pixels (the values are $2.9 \times 10^{-7}$ and $3.72 \times 10^{-10}$ when all pixels are included). These values are smaller than the typical >0.05 value for samples from the same underlying distribution (J. L. Hodges 1958), indicating some statistical distinction between the single-star and binary populations. While this result is promising, Figure 6 also shows that our model's ability to distinguish between single stars and binaries may be due to a high false-positive binary detection rate. While the upper right quadrant of this figure is predominantly populated with binaries, as we expect, the presence of single stars in this region indicates that our methods are prone to picking up false positives. Furthermore, inspection of the model fits shows that the model residuals are inconsistent with clear binary detections, as discussed below.

Figure 7 shows the same ΔBIC of our binary validation sample plotted as a function of binary separation and Kepler magnitude difference $\Delta m_K$ as reported in K. Sullivan et al. (2024). While many of the binaries in our sample have comparable ΔBIC values to the single stars, binaries with the highest ΔBIC lie in regions that we expect to be most sensitive to binaries—that is, binaries with closer separations and intermediate $\Delta m_K$ values. We expect binaries with closer ($\lesssim 0\farcs4$) separations to show stronger signatures of binarity because: (1) their larger radial velocity offset can alter the spectral line shape, even when the features overlap to within one $\sim$10 km s$^{-1}$ resolution element; and (2) the secondary flux is more likely to be entirely captured by the HIRES slit. Furthermore, binaries with intermediate $\Delta m_K$ (i.e., intermediate $T_{\rm eff}$ ratios) will differ the most from similar single-star spectra, based on our simulated data analysis (Section 5.1).

Figure 7 also sheds light on our model's limited completeness. Several binaries with companions within $0\farcs4$ and intermediate $\Delta m_K$ values have $\Delta{\rm BIC} < 300$, indistinguishable from single stars. Furthermore, the fact that many of our single stars have comparable ΔBIC values to the high-ΔBIC binaries in our validation sample suggests that even single-star spectra benefit significantly from the increased complexity of our binary model. In short, our model's limited completeness and high false-positive rate contribute to its inability to identify binaries in our validation sample.

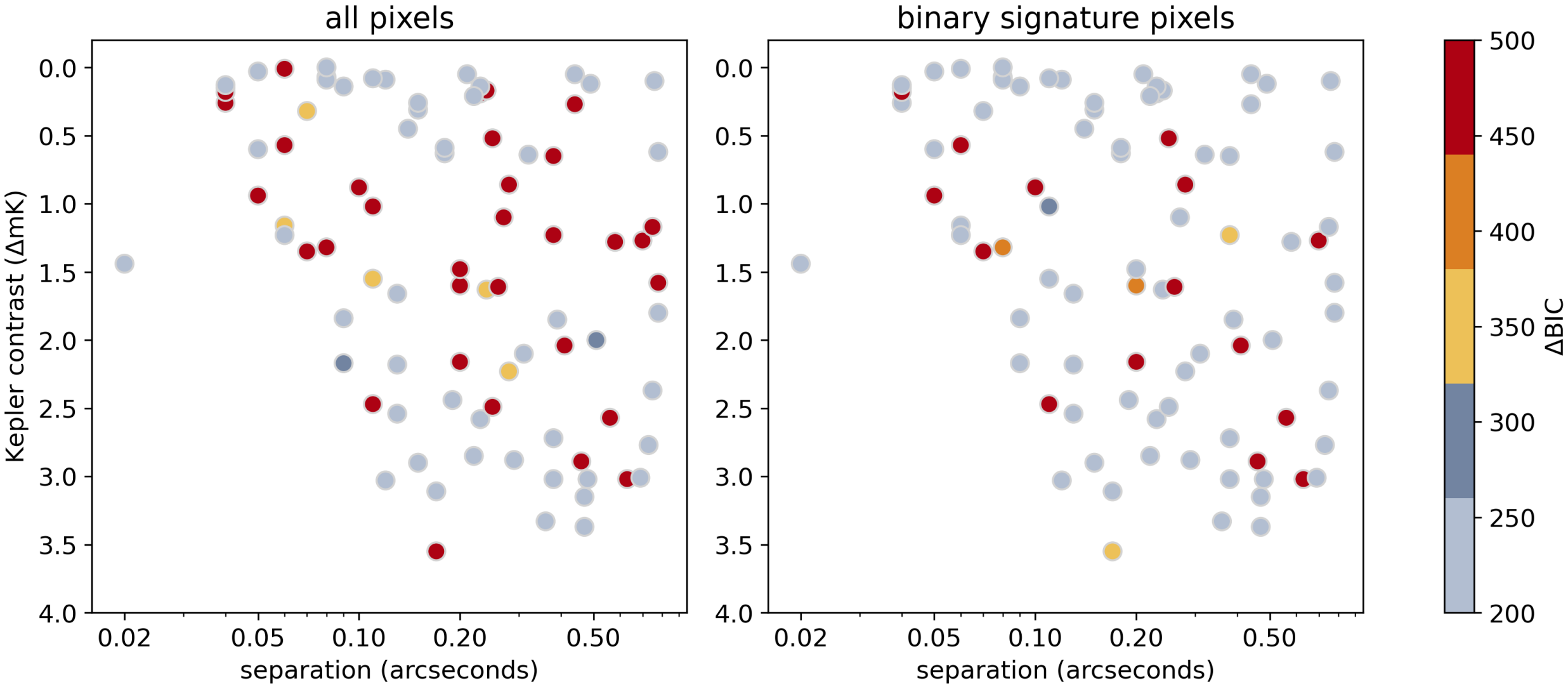

**Figure 7.** Magnitude difference $\Delta m_K$ versus binary separation for all of the binaries in our validation sample, colored according to $\Delta$BIC. While our model is unable to distinguish between the single stars and binaries in most cases, the binaries with the highest $\Delta$BIC values reside in regions of parameter space that we expect to be most sensitive to binaries.

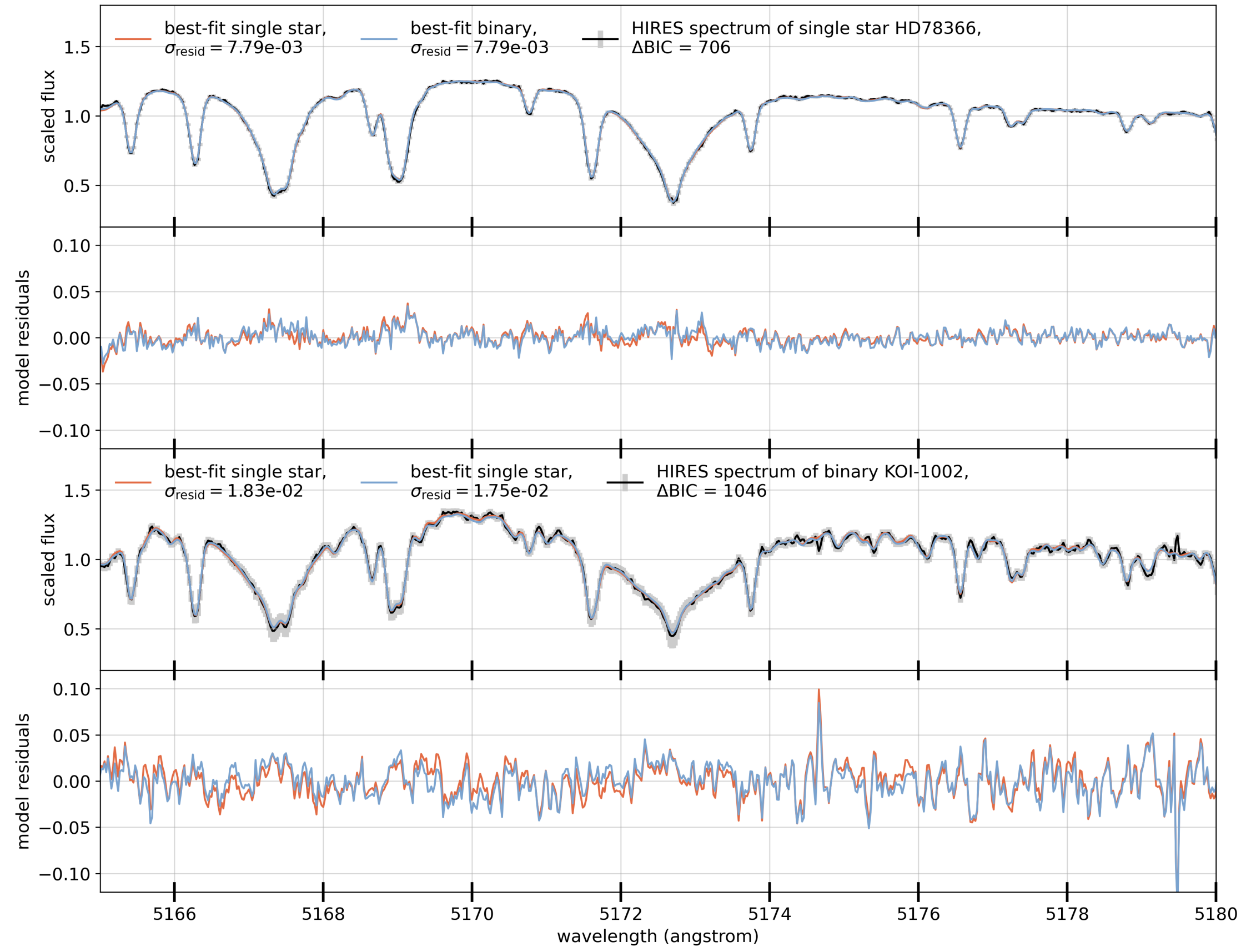

**Figure 8.** HIRES spectra of the single star HD 78366 and binary KOI-1002, with their best-fit single-star and binary spectra and corresponding residuals' standard deviation ($\sigma_{\rm resid}$). We see that in both cases, the single star and binary show $\approx$2%–5% disagreement with the HIRES spectra, limiting our ability to detect signatures of spectral binaries at sufficiently high significance.

Figure 8 shows the best-fit single-star and binary models for two test cases—KOI-1002, a binary with a 5863 K primary and 4714 K secondary that we expect to detect based on ab initio models, and HD 78366, a single star with similar $T_{\rm eff}$ = 5984 K to the primary star of KOI-1002. We also show the standard deviation of the residuals ($\sigma_{\rm resid}$) for each model fit. In both cases, the residuals for the single-star and binary model fits are comparable, with no significant improvement from the binary model. Additionally, the best-fit single-star residuals can be as high as $\approx$2%–5%, even for a single star where we expect the model to agree with the spectrum to within the measurement uncertainties. We interpret this as evidence that our ability to distinguish between single stars and binaries is limited by our model predictions of single-star spectra. If our binary model is only as accurate as its single-star components, then we can only expect to model binary spectra accurately to within 2%–5% of the true binary flux. This is larger than the accuracy we need to detect binaries in most cases, as seen in Section 5.1. Thus, we suspect that our model's per-pixel flux predictions are not currently accurate enough to detect binaries.

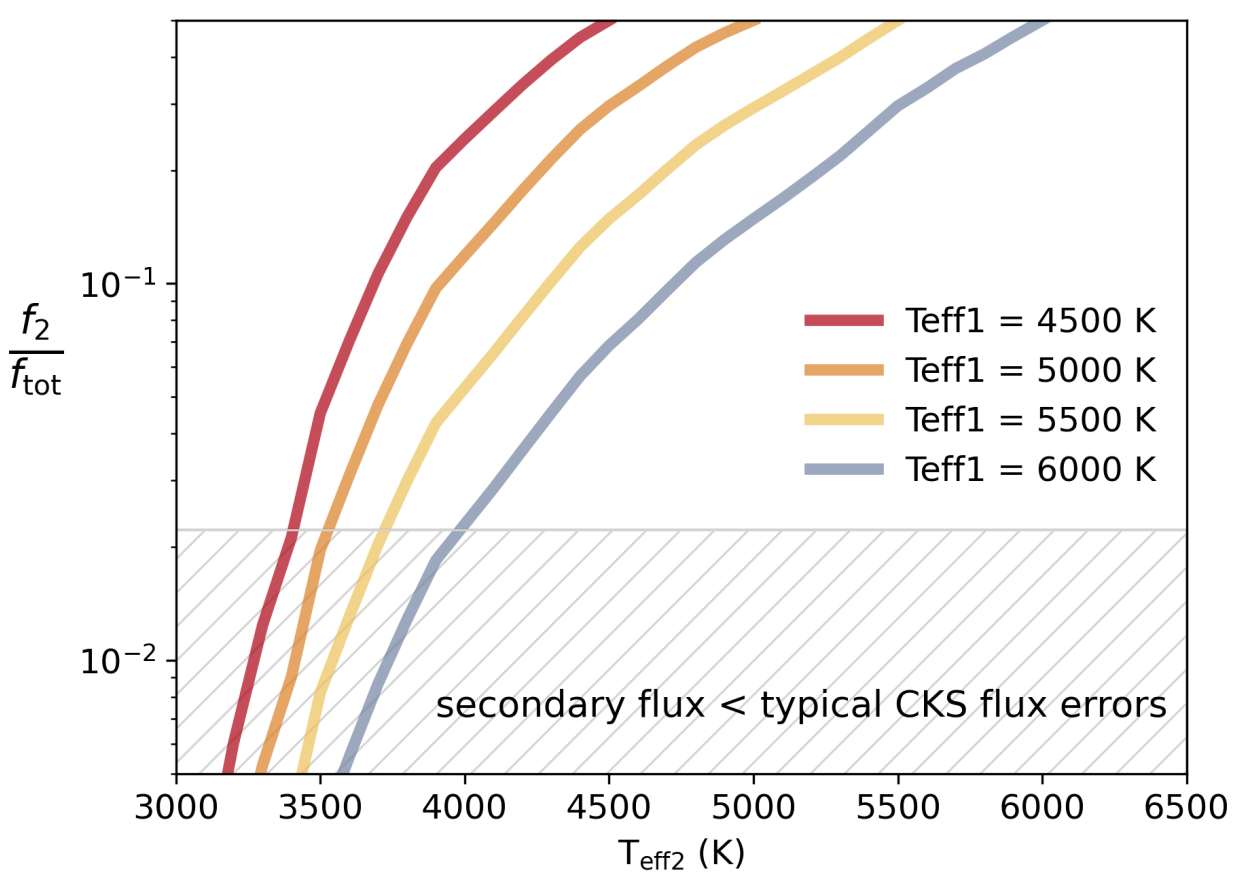

**Figure 9.** Flux contributions from the secondary star evaluated for binaries with a range of primary and secondary temperatures. For a HIRES spectrum with typical S/N = 45 and 2.2% flux errors, secondary stars as cool as ∼3500 K should contribute to flux above the photon noise.

## 5. Discussion

### 5.1. Context from Simulated Data

In general, we expect to be sensitive to any spectral binary whose secondary star contributes more flux than the per-pixel flux uncertainties associated with the spectrum. In practice, binaries are typically detectable when the secondary spectrum deviates enough from the primary to produce unique markers of a nonsingle star at sufficiently high significance on the composite spectrum (K. El-Badry et al. 2018b; M. Kovalev & I. Straumit 2022; M. Kovalev et al. 2024). However, a secondary star that is too similar in spectral type to its primary may not produce such markers, since the composite spectrum will closely resemble that of the primary alone. To get a sense of our expected sensitivity to different types of binaries, we calculated the fractional flux contributions of the secondary star for a grid of binaries with varying primary and secondary $T_{eff}$:

$$\frac{f_2}{f_{tot}} = \frac{1}{1 + f_{rel}}. \quad (16)$$

Figure 9 shows $f_2/f_{tot}$ as a function of secondary temperature $T_{eff2}$ for a range of primary star temperatures $T_{eff1}$. We label $f_2/f_{tot} = 0.022$ for reference, since these are the flux errors associated with a typical CKS spectrum with S/N = 45. We consider spectraS/N below this to be unusable for our purposes, since the secondary flux would be lost to photon noise. Based on this figure, for a median-temperature CKS star with $T_{eff} = 5500$ K, we expect to be sensitive to secondary stars with $T_{eff2} \gtrsim 3700$ K.

We note that while this is a compelling proof of concept, this picture is complicated by two additional factors: first, binary spectra where the primary and secondary are spectrally similar (i.e., residing near the top of Figure 9) may not result in exhibiting significant markers of binarity, compromising our sensitivity to more massive secondaries; and second, the relationship between the secondary and primary star might deviate from the simple temperature–flux relationship we assume for this calculation—in reality, the full spectrum contains pixels with different levels of information and photon noise that depend on both stellar properties and instrumental effects. To get a more accurate sense of the types of binaries we can detect, we developed a model that predicts the HIRES spectra of spectral binaries based on synthetic HIRES-like spectra. To do this, we used Starfish (I. Czekala et al. 2015) to simulate a grid of PHOENIX library spectra (T. O. Husser et al. 2013) for $T_{eff}$ = 3000–7000 K, convolved to the HIRES instrumental profile (FWHM = 4.3 km s$^{-1}$) and resampled to the wavelengths of the HIRES middle-detector CCD chip ($\lambda = 4985-6410$ Å). We removed the continuum by dividing each individual spectrum order by a line fit to the continuum pixels (identified as pixels within 1% of the median flux) of that order.

To predict the HIRES flux of a single star, $f(T_{eff1})$, we linearly interpolated between the two nearest grid spectra, weighted according to their distance from the $T_{eff}$ we are modeling. We then modeled the HIRES flux of a spectral binary $f_{binary}(T_{eff1}, T_{eff2})$ as a linear combination of the primary and secondary flux:

$$f_{binary} = W_1 \times f(T_{eff1}) + W_2 \times f(T_{eff2}), \quad (17)$$

where $W1$ and $W2$ are estimated from the same empirical magnitude–temperature relations described in Section 2.4.

We computed the best-fit single-star and binary models for a simulated binary spectrum with `lmfit.minimize()`, which uses linear least-squares to determine the $T_{eff}$ and ($T_{eff1}$, $T_{eff2}$) that minimize the $\chi^2$ of the single-star and binary models, respectively. We then evaluated the ΔBIC of the best-fit single-star and binary models to determine the extent to which our ab initio spectra were better described by our spectral binary model.

Figure 10 shows an example of a simulated binary spectrum with $T_{eff1} = 5900$ K, $T_{eff2} = 4700$ K, and its best-fit single-star and binary models. In the top panel, we plot the individual primary and secondary star components, along with their composite spectral binary that we fit to. Below it, we plot the spectral binary degraded to S/N = 500, along with the best-fit single-star and binary models and their residuals, to demonstrate our ability to distinguish between the single-star and binary scenarios in the limit of high S/N. The bottom panel shows the same thing, but the spectrum is degraded to S/N = 45, to more closely resemble the CKS spectra. As we see from this figure, for a Sunlike primary and 4700 K secondary, the ΔBIC strongly favors the binary even in the lower-S/N regime.

Next, we simulated $f_{binary}(T_{eff1}, T_{eff2})$ for a range of $T_{eff1}$ and $T_{eff2}$ values and injected Gaussian noise into the spectra to degrade them to S/N = 45. We then computed their best-fit single-star and binary models and their associated ΔBIC values. The ΔBIC values are plotted as a function of the simulated binary $f_2/f_1$ for a number of different $T_{eff1}$ values in Figure 11. Based on this figure, we expect to be sensitive to binaries with $f_2/f_1 \sim$ 0.1–0.6, with increased sensitivity to binaries with cooler primaries where the secondary contributes a larger fraction of the total flux. This is likely because cooler stars have deeper lines that are more sensitive to changes in temperature and therefore more sensitive to the addition of a second star with a slightly different temperature. Additionally, the secondary star will contribute to a larger fraction of the total flux for fainter primaries. Our sensitivity drops off at low $f_2/f_1$, where the secondary is too faint, and high $f_2/f_1$, where the normalized composite spectrum too closely resembles the spectrum of the primary alone.

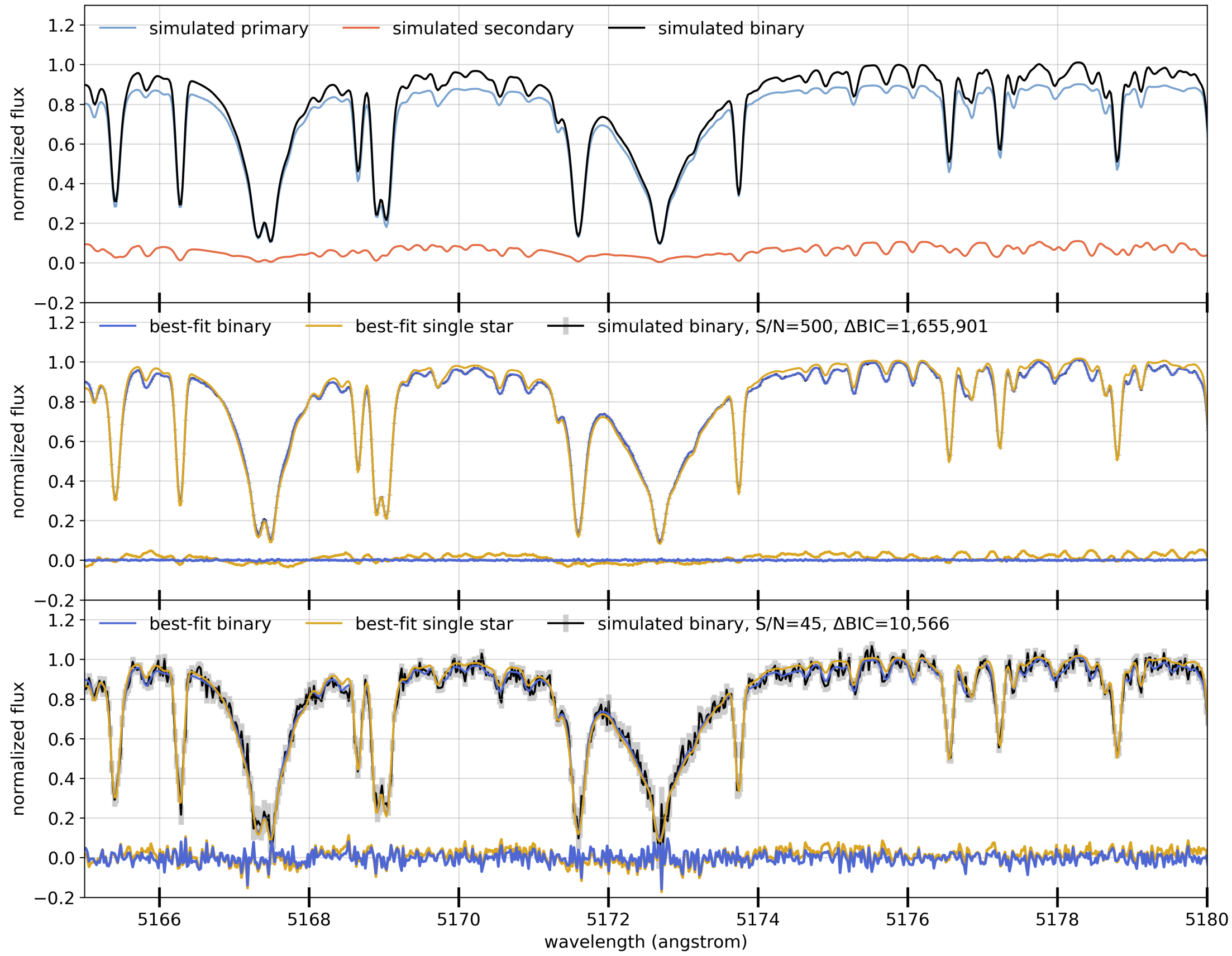

**Figure 10.** Top: simulated spectra of a 5900 K primary (blue) and 4700 K secondary (red) and their composite spectrum (black), weighted according to their relative flux contributions. Middle: composite spectrum with added Gaussian noise and per-pixel S/N = 500, with best-fit single-star and binary models and their corresponding residuals overplotted. The binary fit is strongly favored over the single-star fit with $\Delta\mathrm{BIC} = 1.7 \times 10^6$. Bottom: the same as the middle panel but with a realistic HIRES S/N of 45. The binary fit is still favored over the single star, with $\Delta\mathrm{BIC} = 10^4$.

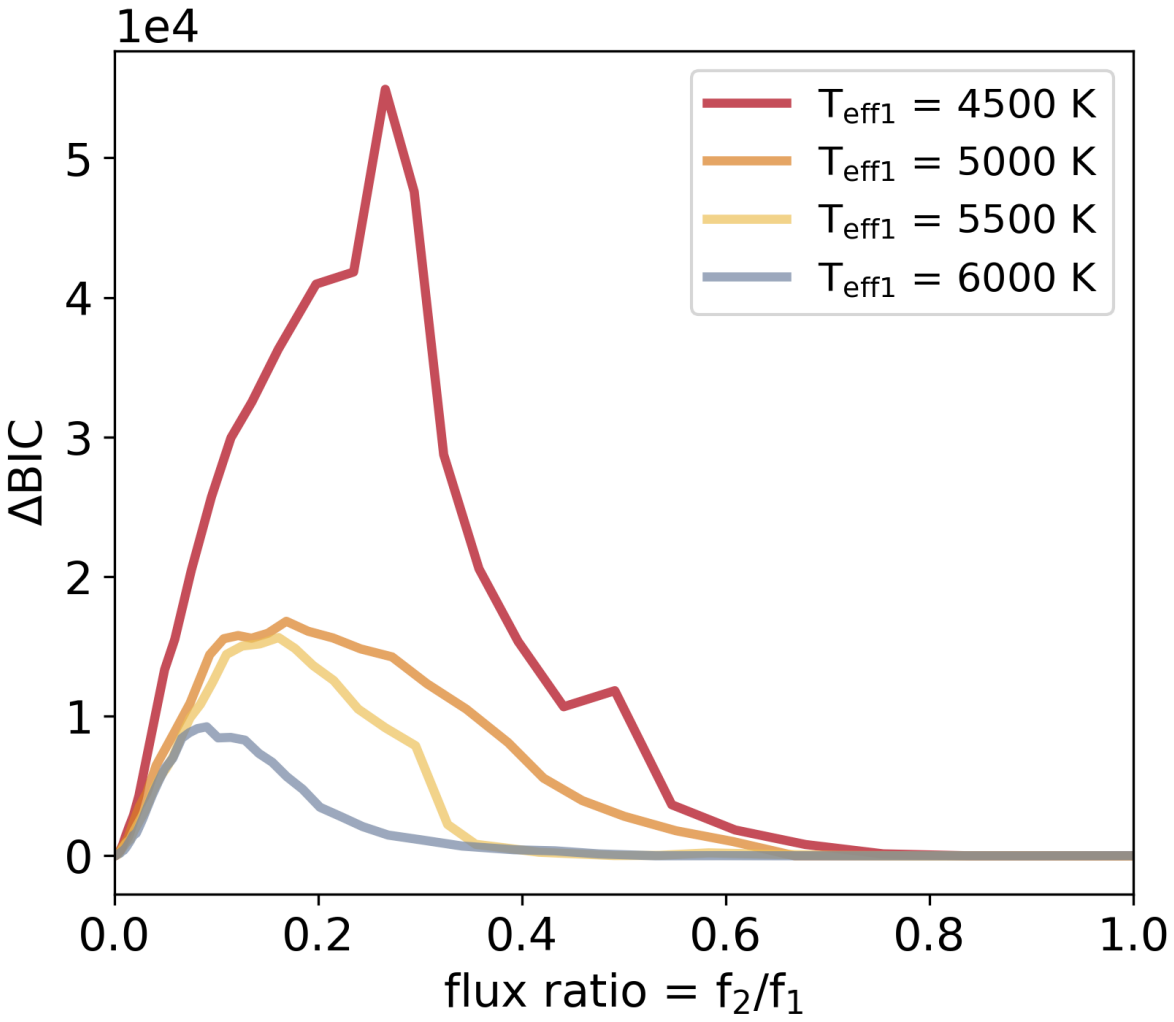

**Figure 11.** The difference in BIC between the best-fit single-star model and best-fit binary model is plotted versus the secondary-to-primary flux ratio for simulated binary spectra with a range of primary star temperatures. The simulated spectra include realistic CKS spectral resolutions and S/Ns. The binary scenario is strongly favored over the single-star scenario for binaries with $f_2/f_1 \approx 0.1$–0.6, particularly for binaries with cooler primaries.

This ab initio experiment makes a number of simplifications to the data-driven methods used in the analysis of this paper. First, our model only varies the $T_{\mathrm{eff}}$ of the primary and secondary stars, while other important labels are fixed to $\log g = 4.5$ dex, $[\mathrm{Fe/H}] = 0$ dex, and $v \sin i = 0\ \mathrm{km\ s^{-1}}$. This is a simplification of the real spectra, which contain single stars and binaries spanning a wide range of stellar label space. Additionally, the PHOENIX models are not a perfect representation of the HIRES spectra we use in our analysis, and best-fit ab initio spectra can disagree with the observed HIRES flux by as much as $\sim$20%. Still, this analysis serves as a compelling proof of concept that suggests that signatures of spectral binaries may be present and detectable in HIRES data.

### 5.2. Future Work

In this paper, we trained a data-driven model of HIRES spectra using the `Specmatch-Emp` stellar library as our training data. We demonstrate that our model is competitive with existing models (S. W. Yee et al. 2017; M. Rice & J. M. Brewer 2020) at deriving stellar labels that agree with high-fidelity labels of previous surveys. We also develop a prescriptive, repeatable wavelet-filtering method for removing low-frequency nonastrophysical variations in our spectra, mitigating night-to-night variations in stellar spectra due to systematics. This processing step significantly improves the accuracy of our model's stellar label predictions, particularly for low-S/N spectra, where traditional continuum treatments are less effective.

We used our data-driven model of single stars to develop a data-driven emulator of spectral binaries. Experiments with synthetic HIRES spectra in this analysis suggest that accurate models of HIRES spectra can discern between spectra of single stars and spectra with signatures of stars with unresolved stellar companions. We find that while our single-star model is able to accurately determine stellar labels, its predictions of the per-pixel HIRES flux are around ∼2%. This limits the degree to which our spectral binary emulator can model the spectra of binaries in our validation sample, which, consequently, limits our model's ability to identify subtle (∼2%) deviations from single-star behavior that our binary spectra should exhibit.

It is also possible that the wavelet-based filtering methods outlined in Section 2.2.3 remove spectral information that is needed to distinguish between single-star spectra and spectral binaries. While we have demonstrated that this loss of information is not prohibitive for performing precise label transfer, it may hinder analysis that depends more sensitively on a small number of deep spectral features, which are more susceptible to alterations during filtering. It also may contribute to our inability to identify spectral binaries, whose deviations from single-star behavior are subtle and highly sensitive to the relative depths of spectral features. This is a general concern for any empirical continuum normalization and motivates future work on robust methods for correcting uncertain instrumental effects on the spectrum while preserving these subtle features.

One of our primary conclusions from our analysis is that the accurate emulation of stellar spectra, even in instances of accurate classifications of stellar labels, is challenging. We see that our model is excellent at label transfer but falls short of emulation. We suspect that this is because data-driven models may under- and overestimate the true flux at different pixels; these discrepancies can balance one another out, to provide reliable stellar labels, but will compound to produce a poor overall goodness of fit across the spectrum. Furthermore, the ∼2% accuracy of our model's flux predictions is comparable to the intrinsic scatter of other data-driven models in the literature that are successful at label transfer (see, for example, Figure 5 of M. Ness et al. 2015 and Figure 6 of M. Rice & J. M. Brewer 2020). It is likely that these models may face similar challenges in applications that require accurate spectral emulation.

While K. El-Badry et al. (2018b) successfully identified spectral binaries in the APOGEE data, this analysis demonstrates that applying these methods to different datasets is less straightforward. The HIRES spectra likely have higher latent dimensionality than the APOGEE spectra—beyond the fact that these spectra are not observed with uniform spectral resolution, the fiber scrambler of HIRES has finite stability, and there is inherent scatter in the resolutions of spectra obtained under the same observing conditions (J. F. P. Spronck et al. 2015). This latent dimensionality adds a layer of complexity to our analysis that compromises our ability to train accurate spectral models and distinguish between astrophysical and instrumental signals in our stellar spectra.

Assembling a large homogeneous training set is also challenging for these data. While the `Specmatch-Emp` library provides an ideal training set in many regards, its labels are compiled from a patchwork of surveys with varying precision and underlying biases. This is compounded by the fact that our binary detection performance depends most sensitively on how well The Cannon works in the low-$T_{\rm eff}$ regime (see Figure 11), where `Specmatch-emp` $v \sin i$ information is incomplete. Beyond this, assembling a clean training set of single stars poses a significant challenge in data-driven searches for spectral binaries. Indeed, many apparently single APOGEE stars reported in K. El-Badry et al. (2018b) were later identified as SB2 candidates in LAMOST-MRS data (M. Kovalev & I. Straumit 2022). Due to these compounding factors, a clean training set composed of thousands of single stars with uniformly derived labels spanning such a wide temperature range is not possible with the existing HIRES data.

It is worth exploring the benefits of utilizing data-driven models with higher complexity for the purposes of accurate spectral emulation. Here, we demonstrate that a piecewise Cannon model is an effective means of adapting quadratic-in-labels models for training sets that span a wide range of stellar temperatures. However, higher-complexity models may be a more suitable solution. For example, the neural network framework of The Payne (Y.-S. Ting et al. 2019) models much more complex flux–label relationships in its training data and has been shown to work with both real and synthetic training data (see, for example, K. El-Badry et al. 2018b; Y.-S. Ting et al. 2019); however, we saw that in our case this model was prone to overfitting to training spectra. In the future, it will be worth exploring The Payne's applications to HIRES spectra further, along with other higher-complexity models, such as higher-order polynomials or modeling a Gaussian process at every pixel, as proposed in M. Ness et al. (2015). It will also be worth investigating the ways in which the intrinsic model scatter can be incorporated into model comparisons. In particular, assumptions encoded in the ΔBIC metric we used may break down in cases of large model uncertainty (i.e., large $s_j^2$).

Looking to the future, there remain opportunities for models with more detailed spectral emulation enabling the detection of unresolved planet-hosting binaries. While we have seen that large surveys like APOGEE and LAMOST provide ideal training sets for data-driven models, it is likely that applications to search for binaries in the Kepler and TESS fields will require data-driven models of spectra from dedicated ground-based surveys to achieve the S/Ns needed for more distant planet hosts at ∼1 pc. Beyond this, processing pipelines of large surveys may produce spectra that are less ideal for spectral binary detection, as was shown to be the case for Gaia DR3 spectra (I. Angelo et al. 2024). More promising are high-resolution spectrometers like HARPS-N and the Keck Planet Finder (KPF), whose stable PSF and legacy data may offer an opportunity to assemble large homogeneous training sets with high-complexity models. These may bring us closer to detailed and accurate spectral emulation—an enormous achievement, with far-reaching applications across all subfields of astronomy.

## Acknowledgments

I.A. is supported by the Michael A. Jura Fellowship at UCLA. I.A. would also like to thank Sam Yee, Adam Kraus, Jack Lubin, Smadar Naoz, Kareem El-Badry, Greg Gilbert, Judah Van Zandt, David Ciardi, Brad Hansen, and Tuan Do for their helpful insights on the analysis portion of this paper.

*Software*: astropy (Astropy Collaboration et al. 2013, 2018), numpy (S. Van Der Walt et al. 2011), matplotlib (J. D. Hunter 2007), scipy (E. Jones et al. 2001), pandas (W. McKinney 2010), The Cannon (A. R. Casey et al. 2016).

## ORCID iDs

Isabel Angelo https://orcid.org/0000-0002-9751-2664
Erik Petigura https://orcid.org/0000-0003-0967-2893
Megan Bedell https://orcid.org/0000-0001-9907-7742